# Image analysis of polycrystalline solar cells and modeling of intergranular and transgranular cracking

Andrea Infuso[1], Mauro Corrado[1], Marco Paggi[2*]

[1]Politecnico di Torino
Department of Structural, Geotechnical and Building Engineering
Corso Duca degli Abruzzi 24, 10129 Torino, Italy

[2]IMT Institute for Advanced Studies Lucca
Piazza San Francesco 19, 55100 Lucca, Italy

**Abstract**

An innovative image analysis technique is proposed to process real solar cell pictures, identify grains and grain boundaries in polycrystalline Silicon, and finally generate finite element meshes. Using a modified intrinsic cohesive zone model approach to avoid mesh dependency, nonlinear finite element simulations show how grain boundaries and Silicon bulk properties influence the crack pattern. Numerical results demonstrate a prevalence of transgranular over intergranular cracking for similar interface fracture properties of grains and grain boundaries, in general agreement with the experimental observation.

## 1. Introduction

Photovoltaics (PVs) based on Silicon semiconductors is the most growing technology in the World for renewable, sustainable, non-polluting, widely available clean energy sources. Commercial PV modules are composite laminates with very different layer thicknesses. Thin Silicon cells are usually embedded in an encapsulating polymer layer (EVA) covered by a much thicker tempered glass [1] (see Fig. 1(a)). In other cases, symmetric glass-polymer-

---

[*] Corresponding Author (Marco Paggi). E-mail: marco.paggi@imtlucca.it, Tel: +39-0583-4326-604, Fax: +39-0583-4326-565

Silicon-polymer-glass laminates are used, especially for semi-transparent facades (see Fig. 1(b)).

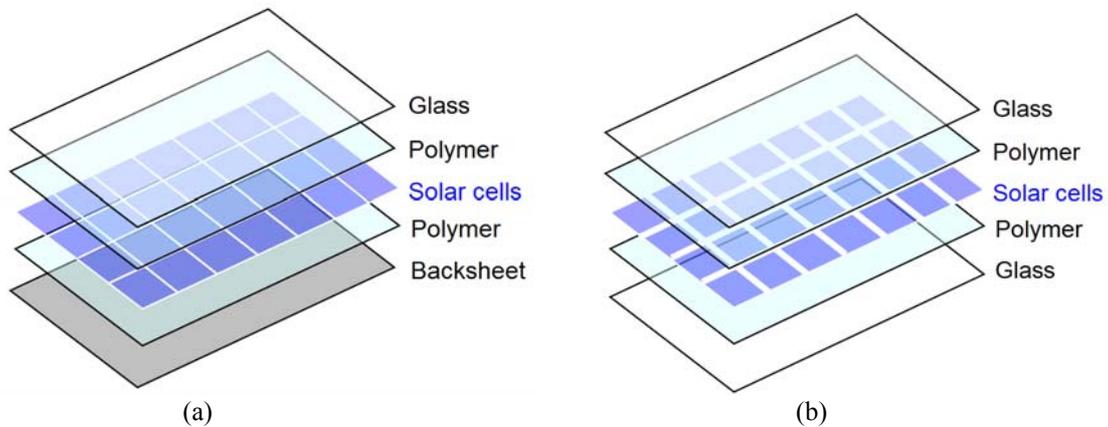

Figure 1. Two typical stacks of PV modules: (a) PV module for roofs or for solar fields; (b) PV module for semi-transparent roofs or facades.

The majority of solar cells available on the market are made of either monocrystalline or polycrystalline Silicon. Solar cells are separated in their plane by a variable content of EVA, depending on the amount of shading requested. Two main semiconductors, called *busbars*, electrically connect the cells in series. Very thin Aluminium conductors perpendicular to the busbars, called *fingers*, are also present to collect the electrons originated by the photovoltaic effect from the surface of the cells to the busbars (see Fig. 2).

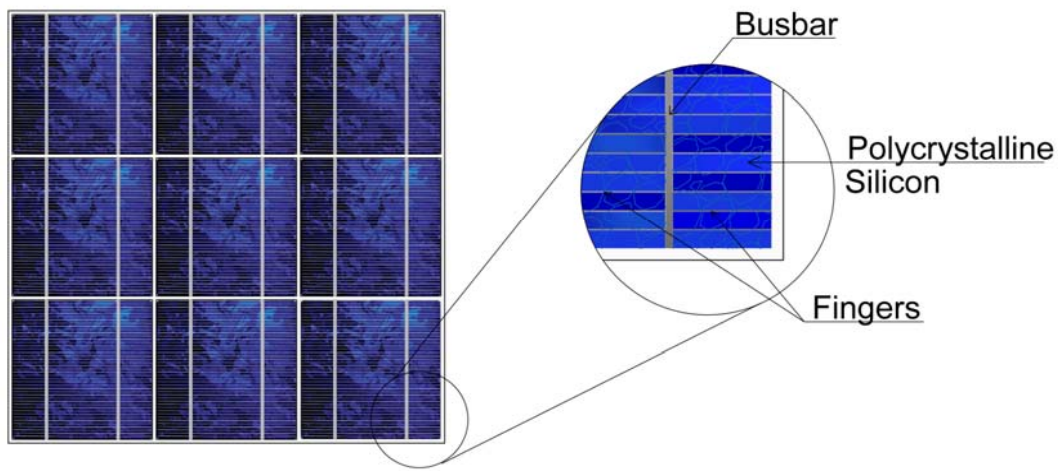

Figure 2. Sketch of a PV module and detail of busbars and fingers over the Silicon cell

The quality control of these composites is of primary concern from the industrial point of view. On the one hand, the aim is to develop new manufacturing processes able to reduce the number of cells or modules rejected. On the other hand, even if all the damaged cells are theoretically discarded during manufacturing, it is impossible to avoid the occurrence of microcracking during the subsequent stages. Sources of damage in Silicon cells are transport, installation and use (in particular impacts, vibrations, snow loads and environmental aging caused by temperature and relative humidity variations) [2-4]. Since microcracking can lead to large electrically disconnected areas, there is an urgent need to understand the origin of this phenomenon and find new technical solutions to improve the durability of PV modules.

To investigate the effect of mechanical loads on cracking in solar cells, mini-modules of 10 cells disposed along two rows (5 cells per row) have been subjected to 4-point bending in [5] (see Fig. 3(a)). The force-displacement curve obtained from this test is depicted in Fig. 3(b) and shows brittle failure as soon as cracking propagates. Microcrack patterns, impossible to be detected by a naked-eye inspection of the cells, were monitored by using the electroluminescence (EL) technique, see Fig. 4 [5]. These images refer to the portion of the module span where the bending moment is constant. Cracks develop along some preferential lines almost parallel to the direction of line loading. In case of horizontal busbars perpendicular to the line of loading, Fig. 4(a), a diffuse crack pattern is observed with the appearance of crack branching. For vertical busbars parallel to the line of loading, Fig. 4(b), single cracks propagate and lead to large electrically disconnected black areas. The orientation of busbars and of the thin electric fingers with respect to the direction of application of loading has therefore a role on the crack pattern at failure [5]. Both transgranular and intergranular cracks are clearly present and should be considered in numerical models, although a qualitative visual inspection of Fig. 4 would suggest that transgranular cracking is more frequent than the intergranular one.

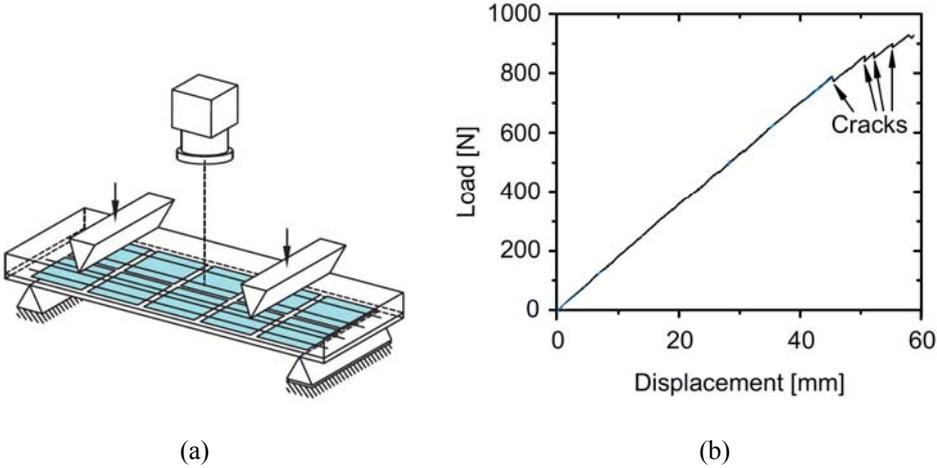

(a)          (b)

Figure 3. Setup of the experimental test on mini-modules carried out in [5] (a) and obtained load-displacement curve (b) (adapted from [5]).

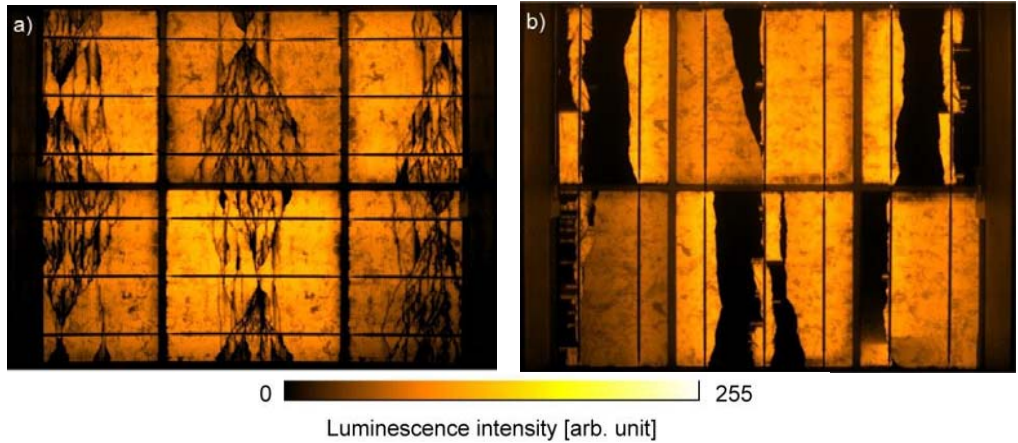

Figure 4. Crack patterns of mini-modules made of polycrystalline Silicon tested according to the setup in Fig. 3 (adapted from [5]). Line loading perpendicular to busbars (a); line loading parallel to busbars (b).

From the modelling point of view, a multi-physics and multi-scale computational model has been proposed in [6]. The original idea is to couple elastic, thermal and electric fields to achieve a predictive stage in the numerical simulations. To simulate fracture in solar cells of a commercial PV module, structural analysis has been performed by using the finite element method and by considering the laminate as a multi-layered plate. The computed in-plane displacements at the boundaries of the cells were transferred to the micro-models of the individual cells, where the actual material microstructure was considered. In [6], intergranular cracking was considered as the only possible source of damage. Further progress has been presented in [7,8], where coupling between the elastic and the thermal fields has been accounted for by developing a specific thermo-elastic Cohesive Zone Model (CZM).

In the present study, transgranular cracking, i.e., cracking through the grains, is also considered in addition to the intergranular one, i.e., cracking along grain boundaries, and their competition is examined. To this aim, a novel image analysis procedure for the identification of grains and grain boundaries is proposed. Exploiting a modified intrinsic approach to cohesive crack propagation to reduce mesh dependency, a Matlab pre-processor has been coded to automatically generate finite element (FE) meshes with cohesive interface elements inserted around all the FE edges. Different fracture parameters can be associated to the interface elements depending on their position (along the grain boundaries or inside the grains). Numerical examples are provided to show the applicability of this computational approach to polycrystalline Silicon solar cells. Depending on the cohesive properties associated to the grain boundary and grain interior cracks, the computed crack patterns are compared and general trends are determined.

## 2. Image analysis of polycrystalline Silicon cells and finite element mesh generation

*2.1 Image analysis of solar cells*

The majority of produced solar cells are made of either mono or polycrystalline Silicon. The material microstructure has a role on the electric performance of the cell, in fact grain boundaries are defects introducing additional resistances to the current flow not present in monocrystalline Silicon (see the comparison between electroluminescence (EL) images taken in our laboratory and shown in Fig. 5).

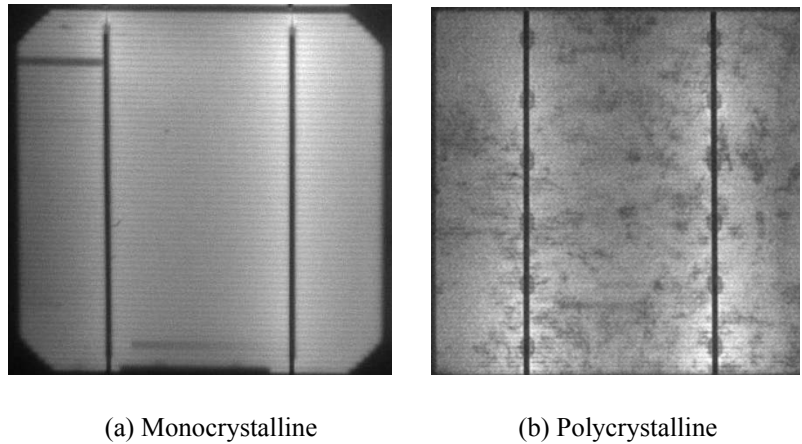

(a) Monocrystalline                     (b) Polycrystalline

Figure 5. EL images of a mono (a) and a polycrystalline (b) solar cell. Grain boundaries lead to a nonuniform EL intensity due to local additional resistances.

The material microstructure plays also a role as far as the mechanical response is concerned. Different grain orientations may lead to different elastic parameters, as usually observed in polycrystals. Moreover, the role of grain boundaries on cracking is an aspect not yet clarified in the literature. To investigate on this issue, the heterogeneity in the material microstructure has to be taken into account in numerical simulations and the proposed models have to be as close as possible to reality. When examining polycrystalline solar cells embedded in photovoltaic modules in daylight conditions, grain boundaries are nearly impossible to be distinguished due to a blue anti-reflective coating which covers the material microstructure (see Fig. 6(a)). Post-processing of EL images may partially help to recognize the grain microstructure by the non-uniform EL emission (see Fig. 6(b)), although the presence of thin electric conductors (fingers) on the cell surface makes grain boundary recognition a not trivial task.

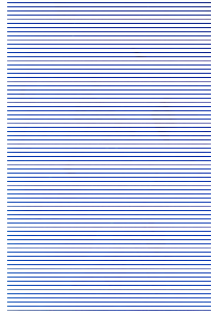 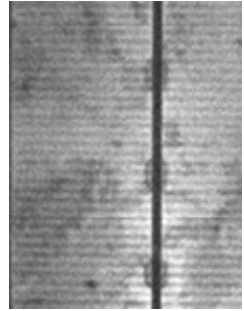

(a) Daylight picture of a cell        (b) EL image of a cell

Figure 6. Grain boundary identification is a challenging task in polycrystalline cells: disturbance of the blue anti-reflective coating in daylight pictures (a); disturbance of fingers (thin horizontal dark lines) and busbars (thick black vertical line) in EL images.

For quantitative image processing, good candidates are Silicon cells without anti-reflective coating used to produce semi-transparent glass-polymer-Silicon-polymer-glass modules for car sheds or facades (see Fig. 7), where fingers are replaced by a thin layer of conducting Aluminium deposited over the whole cell. If illuminated by the sun on one side, the transparency allows material microstructure identification by taking a daylight photo from the opposite side. The only remaining disturbance is given by the two vertical busbars (see Fig. 8). Grains appear of different colours and transparency due to their different orientation. The smallest grain is of a few millimetre size (the lateral size of the cell is about 10 cm). The complexity regards the grain geometry, since very often grains have a very elongated shape due to Silicon wafer processing (see Fig. 8), which poses challenges for their identification and finite element meshing.

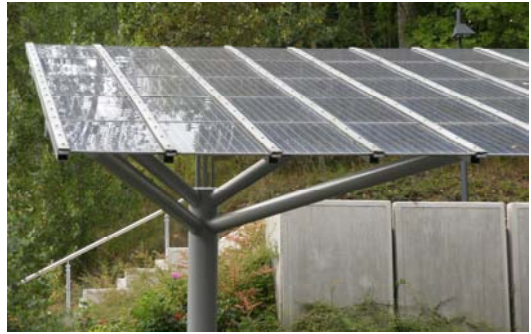

Figure 7. Example of glass-polymer-Silicon-polymer-glass solar modules used for semi-transparent car sheds or facades.

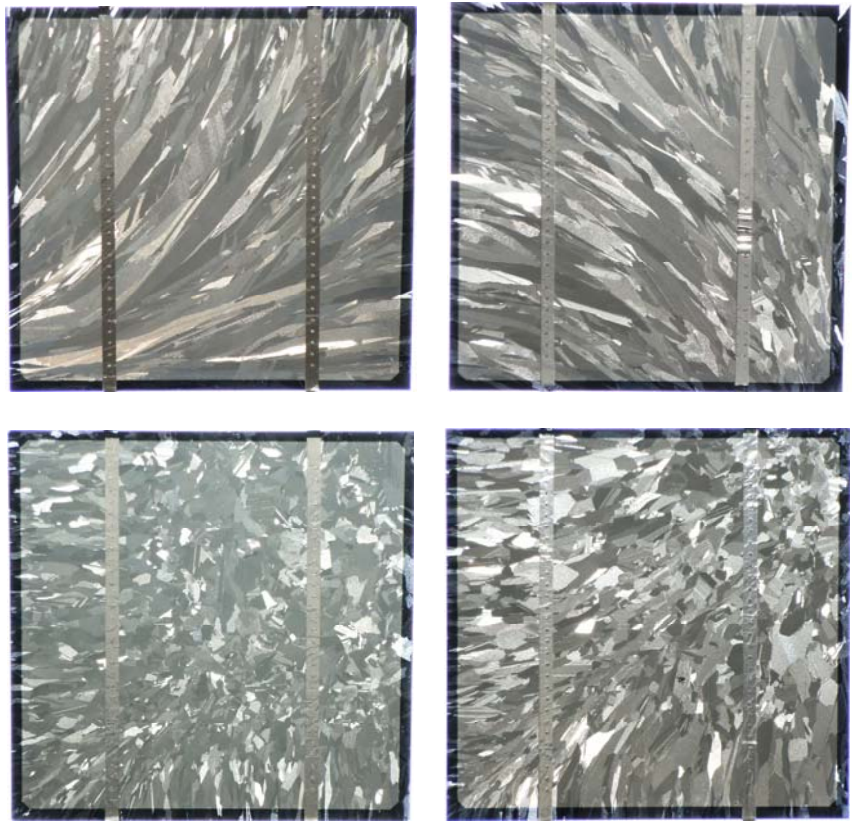

Figure 8. Daylight images of solar cells embedded in glass-polymer-Silicon-polymer-glass photovoltaic modules.

An innovative image analysis procedure is proposed here for grain boundary identification and the subsequent finite element mesh generation. The basic steps are illustrated in reference to Figs. 9 and 10. A high resolution colour image of a cell is the starting point (Fig. 9). The first step is to transform the master colour picture in a grey scale image (Fig. 10(a)). This procedure filters the number of grains in the model by limiting the number of possible orientations to a maximum of 256 (each grain orientation corresponds to a grey intensity varying from 0 to 255). The resulting image is still very complex, with very elongated grains having similar grey levels nearly impossible to be distinguished. It is therefore recommended to further process Fig. 10(a) by transforming it in a black and white image (Fig. 10(b)). This procedure further filters grains close to each other and having very similar orientations. From this operation, however, a noise represented by small black dots is artificially introduced and it is a source of disturbance for the next stage of grain boundary identification. It is therefore useful to apply to Fig. 10(b) an averaging filter over 64 neighbouring points for each image pixel, with a weight of 1/64 for all the pixels. This leads to Fig. 10(c), where black dots are basically removed. Finally, the last step is to identify the grain boundaries. By treating the image as a 3D array with 0 or 1 integer values for each (x,y) pixel position, a point belonging to a grain boundary is identified by scanning the whole image by horizontal cross-section lines and finding the pixels where there is a transition from 0 to 1 or viceversa (see Fig. 10(d)).

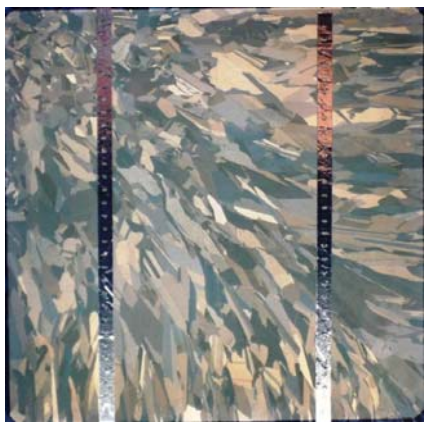

Figure 9. A colour image of a polycrystalline solar cell taken under daylight conditions.

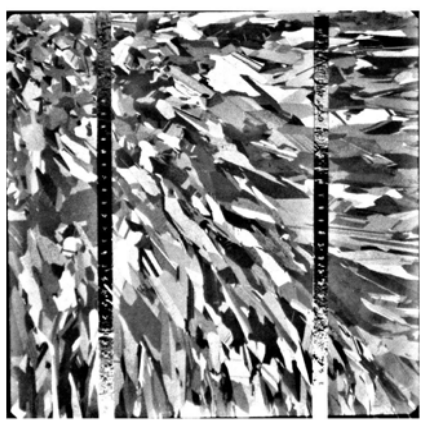

(a) Grey scale image

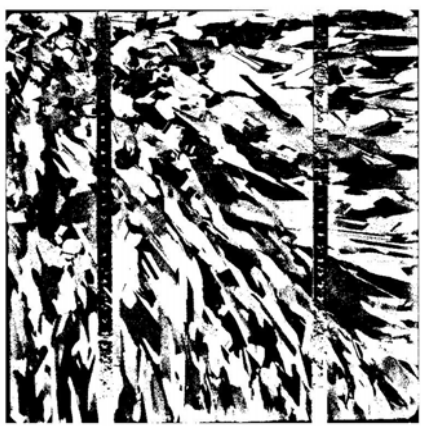

(b) Black and white image

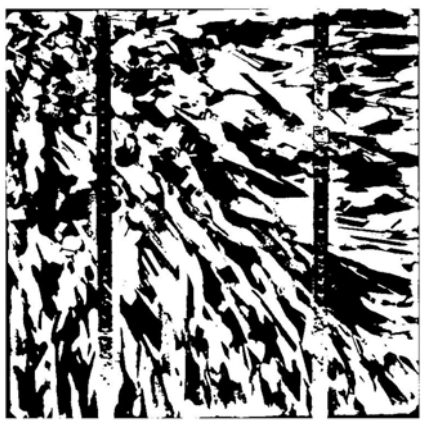

(c) Filtered black and white image

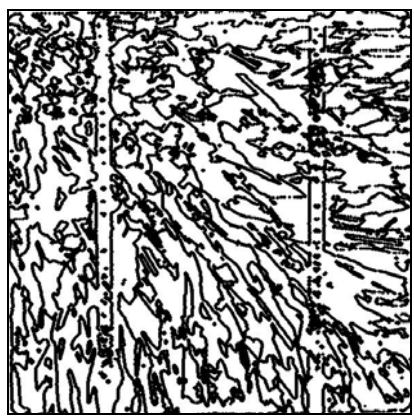

(d) Identified grain boundaries

Figure 10. Some steps of image processing for grain boundary identification.

With the help of Fig. 10(d), grain boundaries are further approximated by polylines as closed polygons whose topological properties (coordinate vertices and connectivity matrix of the edges) are stored in an array for the subsequent finite element mesh generation. The obtained polygons are shown in Fig. 11, superimposed to the original image.

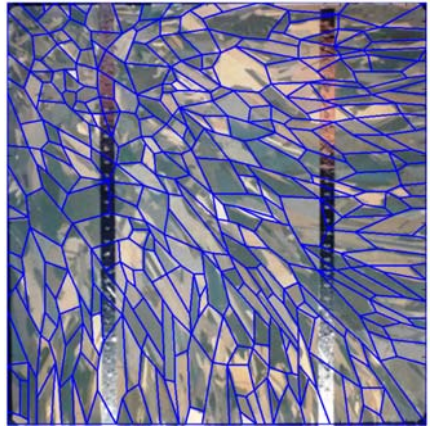

Figure 11. The real image of a polycrystalline Silicon cell with its polygonal grain approximation superimposed.

*2.2 Algorithm for the generation of finite element meshes for fracture mechanics simulations*

The data structure of the various polygons is passed in input to the free software GMSH [9] for meshing the grains as if they were joined. This operation can be done with linear triangular or quadrilateral finite elements with a prescribed mesh density. The output of GMSH is then elaborated by a post-processing code developed in house and written in Matlab. This programme automatically duplicates all the nodes of the finite elements and inserts interface elements all around them. The mechanical properties of the interface elements can be finally attributed depending on their position (inside the grains or along the grain boundaries). The final output is represented by a list of nodal coordinates and by the elements connectivity matrix in a format suitable for running a simulation with the finite element analysis programme FEAP [10]. Other formats used by commercial FE software like Abaqus or Ansys can also be produced.

**3. Constitutive relations, weak form and nonlinear finite element approximation**

For the numerical simulation of transgranular and intergranular cracking in polycrystalline Silicon, a 2D plane stress model is considered. The principle of virtual work reads:

$$\int_V (\nabla \delta \mathbf{u})^T \boldsymbol{\sigma} \mathrm{d}V - \int_S \delta \mathbf{g}^T \mathbf{t} \mathrm{d}S = \int_{\partial V} \delta \mathbf{u}^T \mathbf{f} \mathrm{d}S \qquad (1)$$

where the first term on the left hand side is the classical virtual work of deformation of stresses and strains inside the bulk *V* and the right hand side is the virtual work of tractions acting on the boundaries of the cell ∂*V* for the displacements of their corresponding points of application. The second term on the left hand side is the contribution to the virtual work of the vector of the normal and tangential cohesive tractions **t**=(σ,τ)$^T$ for the corresponding relative opening and sliding displacements **g**=($g_N$,$g_T$)$^T$ at all the element boundaries *S*.

The CZM is adopted to depict the nonlinear process of crack growth in Silicon. According to this approach, tractions normal and tangential to crack faces are functions of the relative opening and sliding displacements of the crack faces themselves. Here, we consider the CZM formulation recently proposed in [7,8], which is based on an analogy between contact mechanics and fracture mechanics and is also suitable for proper modelling the localized additional thermal resistance of cracks in case of coupled thermoelastic analyses. The cohesive tractions are linear increasing functions of the corresponding gaps up to a maximum value reached in correspondence of the dimensionless separation $l_0/R$, where *R* is the root-mean-square roughness of the heights of the crack profile at complete separation. The ratio between the peak cohesive traction and $l_0$ corresponds to the stiffness of the interface element during the linear branch of the constitutive law. Hence, the parameter $l_0$ physically corresponds to the maximum normal gap before reaching the softening in the cohesive zone formulation. Although for adhesive joints it can be related to the actual thickness and elastic properties of the adhesive, in the present study it is mostly a dummy parameter to modify the interface stiffness and avoid mesh-dependent results in the framework of the intrinsic CZM approach. It is therefore tuned by varying the mesh size in order to have the slope of the stress-strain curve of Silicon in the linear stage approximately equal to the Young modulus of Silicon. After the linear branch, an exponential softening is assumed in the CZM formulation. The resulting expressions of the cohesive tractions, accounting for Mode Mixity, i.e., when both opening and sliding take place at the same time, are (see Fig. 12 for a graphical representation):

$$\sigma = \begin{cases} \sigma_{max} \exp\left(\dfrac{-l_0 - |g_T|}{R}\right) \dfrac{g_N}{l_0} & \text{if} \quad 0 \leq \dfrac{g_N}{R} < \dfrac{l_0}{R} \\ \sigma_{max} \exp\left(\dfrac{-g_N - |g_T|}{R}\right) & \text{if} \quad \dfrac{l_0}{R} \leq \dfrac{g_N}{R} < \dfrac{g_{Nc}}{R} \\ 0 & \text{if} \quad \dfrac{g_N}{R} \geq \dfrac{g_{Nc}}{R} \end{cases} \quad (2)$$

$$\tau = \begin{cases} \tau_{max} \exp\left(\dfrac{-l_0 - g_N}{R}\right) \dfrac{g_T}{l_0} & \text{if} \quad 0 \leq \dfrac{|g_T|}{R} < \dfrac{l_0}{R} \\ \tau_{max} \text{sgn}(g_T) \exp\left(\dfrac{-|g_T| - g_N}{R}\right) & \text{if} \quad \dfrac{l_0}{R} \leq \dfrac{|g_T|}{R} < \dfrac{g_{Tc}}{R} \\ 0 & \text{if} \quad \dfrac{|g_T|}{R} \geq \dfrac{g_{Tc}}{R} \end{cases} \quad (3)$$

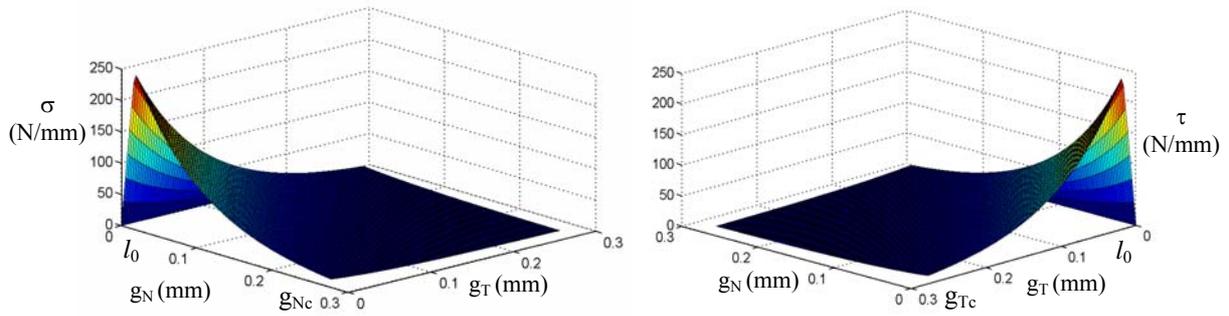

(a) Normal cohesive tractions as a function of crack opening and sliding

(b) Tangential cohesive tractions as a function of crack opening and sliding

Figure 12. Cohesive constitutive laws accounting for Mode Mixity.

The finite element approximation of the weak form in Eq. (1) is achieved by using linear triangular elements for the continuum and linear interface elements for the cohesive cracks. The reader is referred to [11-13] for the matrix formulation, the consistent linearization of the nonlinear CZM for the use of the Newton-Raphson method, and for the numerical implementation in the finite element analysis programme FEAP.

An intrinsic approach is pursued here for simulating crack growth in the material microstructure. Interface elements are inserted all around the triangular elements from the beginning of the simulation. As compared to an extrinsic approach, where interface elements are adaptively inserted in the model at each time step, the advantage is that remeshing operations and nodes and elements renumbering is avoided [14]. On the other hand, the necessity of having an initial linear increasing branch in the CZM to capture the physical condition of vanishing stresses in the bulk in correspondence to vanishing strains, introduces an artificial compliance related to the linear branch of the CZM that may be unrealistic. Another problem is the possible mesh dependency of the crack trajectory, which is constrained by the mesh discretization, and has to be ascertain with care by a suitable mesh dependency study.

In the next section, the issue of mesh dependency of the crack pattern is investigated by considering three different mesh discretizations. To avoid the artificial increase of compliance induced by the intrinsic approach, a modified scheme is proposed where the initial stiffness of the CZM formulation is increased by refining the mesh by suitably reducing the parameter $l_0$ and keeping constant the fracture energy and the peak cohesive tractions (given by the products $\sigma_{max}\exp(-l_0/R)$ and $\tau_{max}\exp(-l_0/R)$), that have to be treated as material properties. The optimal values of $l_0$ to be used in the simulations are chosen so that the initial slope of the obtained homogenized stress-strain curve is almost independent of the mesh size and equal to a value representative of the physical response of a polycrystalline material composing a solar cell, by keeping constant the tensile strength and the fracture energy of Silicon.

It is remarkable to note that this rescaling of the initial compliance of the CZM by keeping constant the fracture energy is not possible with other CZM formulations like that by Tvergaard [15], because it is defined by a single nonlinear function of the gaps over the whole separation range.

## 4. Numerical results

A model problem consisting of 4 grains (lateral size of about 1 cm) is analyzed under plane stress conditions with a typical thickness of solar cells equal to 0.166 mm. Three different FE mesh densities are considered, see Fig. 13. An edge crack is placed on the left side to localize the deformation and simulate the existence of a defect inside a grain. This microstructure is tested under vertical displacement control imposed on the top side to simulate crack propagation in tension. The horizontal side on the bottom is restrained to displacements in the vertical direction, whereas the other sides are free.

The finest mesh size used in this study is expected to provide a reasonable approximation to the stress field according to the criteria proposed in [16-18].

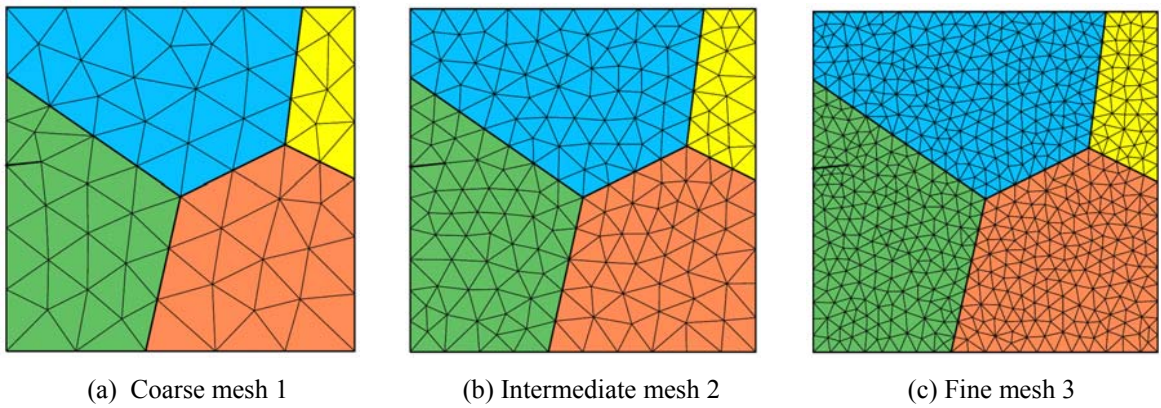

(a) Coarse mesh 1    (b) Intermediate mesh 2    (c) Fine mesh 3

Figure 13. FE meshes with different element size used in the numerical tests. Note the initial crack inserted in the green grain (see the online version of the article for colours).

To investigate on the competition between transgranular and intergranular cracking by varying the fracture properties of the grain boundaries, a parametric study is performed. For the interface elements inside the grains, they are selected to represent the properties of bulk Silicon: $\sigma_{peak} = \tau_{peak} = 1.5$ GPa (to which corresponds a tensile strength of Silicon $\sigma_{peak} = 250$ Nmm/mm$^2$ for a cell with thickness of 0.166 mm), $g_{Nc} = g_{Tc} = 280$ μm. The fracture energy, that for a cell with thickness of 0.166 mm results to be $G_F = 22$ Nmm/mm, is higher than for a stand-alone polycrystalline Silicon and takes into account the toughening effects exerted by the metallic fingers that act as fibers keeping cracks closed and by the viscoelastic epoxy material encapsulating the Silicon cell. The parameter $R$ has been estimated equal to 88.2 μm in order to have the area below the Mode I CZM curve equal to the selected fracture energy [7]. The remaining parameter $l_0$ is tuned to avoid mesh dependency and obtain the same stress-strain curve in the elastic regime regardless of the mesh size: $l_0 = 18.0$ μm for the coarse mesh 1, $l_0 = 10.0$ μm for the intermediate mesh 2 and $l_0 = 5.2$ μm for the fine mesh 3. The CZM parameters of the interface elements along the grain boundaries are modified with respect to those of the bulk by considering $\sigma_{peak}^{GB}/\sigma_{peak}^{grain} = 1.00, 0.75, 0.55, 0.40$, that is, the grain boundary strength is progressively diminished as compared to the bulk strength. All the other fracture parameters are kept constant, so that the fracture energy scales according to the peak strength. Regarding the computational effort, the finest meshes require approximately 1 hour of computing time to determine the whole stress-strain curve until failure.

The computed average axial force per unit length, evaluated by summing up all the nodal reactions in the vertical direction on the top side and dividing the result by the side length, are plotted vs. the imposed strain in Fig. 14. By reducing the ratio between the peak cohesive tractions of the elements belonging to the grain boundary and of those elements inside the grains, $\sigma_{peak}^{GB}/\sigma_{peak}^{grain}$, in the range from 1.00 to 0.40 (from Fig. 14(a) to Fig. 14(d)), grain boundaries become weaker and weaker and the microstructure is more and more prone to intergranular cracking. For each value of $\sigma_{peak}^{GB}/\sigma_{peak}^{grain}$, results obtained from different mesh sizes are almost the same in terms of maximum axial force per unit length. This proves that the procedure adopted to rescale the parameter $l_0$ of the CZM relation to provide a unique initial linear elastic regime in the homogenized curves is effective to obtain mesh-size independent results.

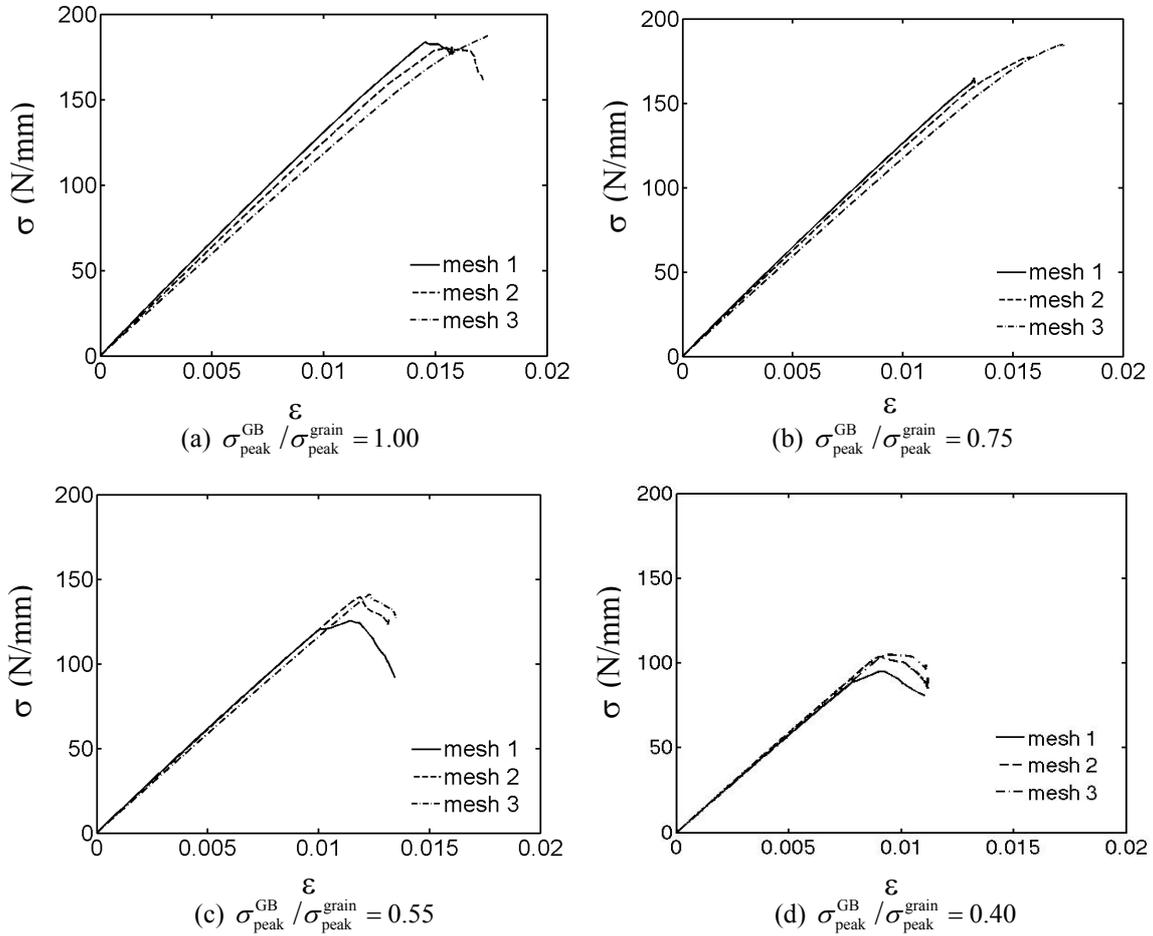

Figure 14. Homogenized mechanical response by varying the ratio between the peak cohesive tractions of the grain boundaries and of the grains, $\sigma_{peak}^{GB}/\sigma_{peak}^{grain}$. In each plot, results for meshes 1-3 in Fig. 13 are compared.

The examination of the contour plot of the vertical displacements at maximum load shows that, if all the interface elements have the same fracture properties regardless of their position (first line in Fig. 15 case (a)), then the initial stress-free defect propagates and transgranular cracking prevails over the intergranular one. Grain boundaries are highlighted by blue lines, whereas the crack path is shown with white lines. By reducing the cohesive peak tractions ratio $\sigma_{peak}^{GB}/\sigma_{peak}^{grain}$ to 0.75, the initial defect propagates inside the grain and then grain boundary decohesion takes place when the defect meets the first grain boundary (second line in Fig. 15 case (b)). For $\sigma_{peak}^{GB}/\sigma_{peak}^{grain} = 0.55$, the initial defect does not propagate any longer and cracking is pure intergranular (third line in Fig. 15 case (c)). This kind of failure is even more evident for lower values, e.g., for $\sigma_{peak}^{GB}/\sigma_{peak}^{grain} = 0.40$, for all the mesh sizes (fourth line in Fig. 15 case (d)).

Correspondingly, the maximum axial force per unit length in the homogenized curves in Fig. 14 depends on the ratio $\sigma_{peak}^{GB}/\sigma_{peak}^{grain}$ and the kind of crack propagation (trans- or intergranular cracking). A decrease in the maximum load of only 6% is obtained by decreasing the ratio $\sigma_{peak}^{GB}/\sigma_{peak}^{grain}$ from 1.00 to 0.75 (case (a) to case (b)). This is due to a prevalence of transgranular cracking, governed by the strength of the bulk material. On the contrary, the reduction of the maximum load from case (c) to case (d) is 25%, almost equal to the decrease in the ratio $\sigma_{peak}^{GB}/\sigma_{peak}^{grain}$. Both of these cases, in fact, are characterized by purely intergranular cracking, governed by the strength of the interfaces. For weaker grain boundaries, the appearance of a softening branch after the peak is observed instead of a catastrophic failure as for transgranular cracking.

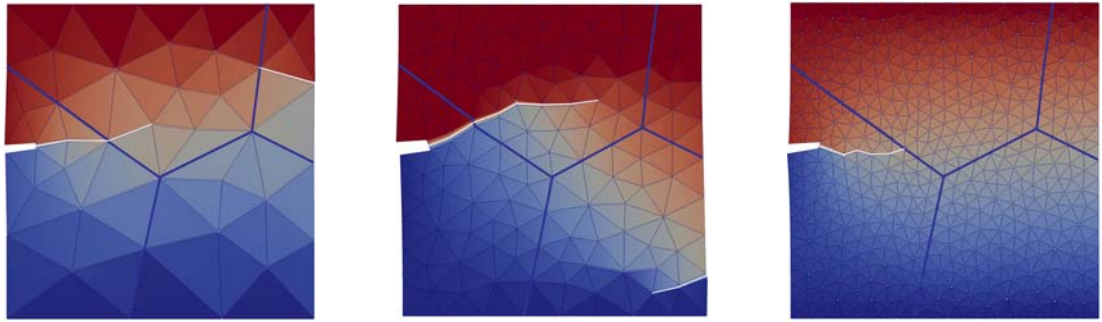
(a) $\sigma_{\text{peak}}^{\text{GB}} / \sigma_{\text{peak}}^{\text{grain}} = 1.00$ and 3 different mesh sizes

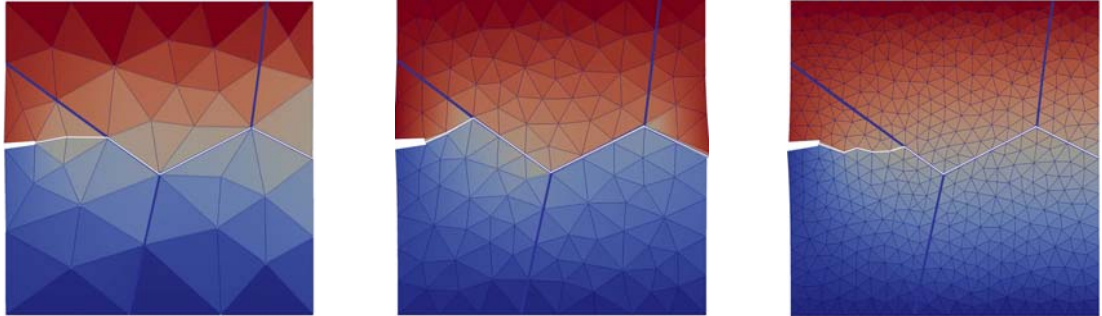
(b) $\sigma_{\text{peak}}^{\text{GB}} / \sigma_{\text{peak}}^{\text{grain}} = 0.75$ and 3 different mesh sizes

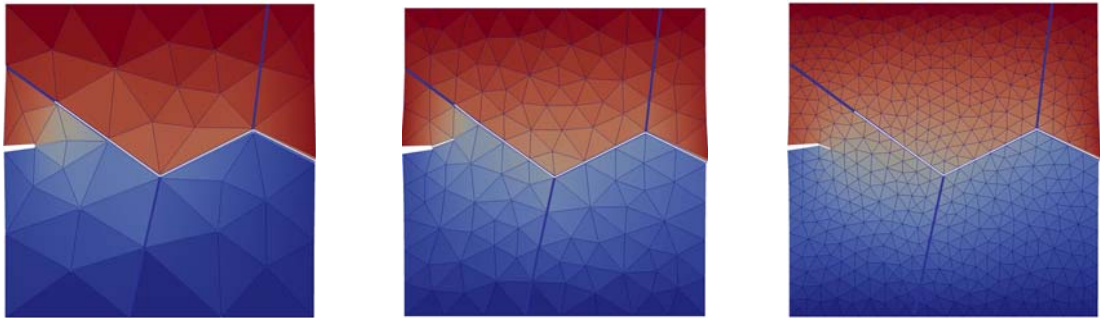
(c) $\sigma_{\text{peak}}^{\text{GB}} / \sigma_{\text{peak}}^{\text{grain}} = 0.55$ and 3 different mesh sizes

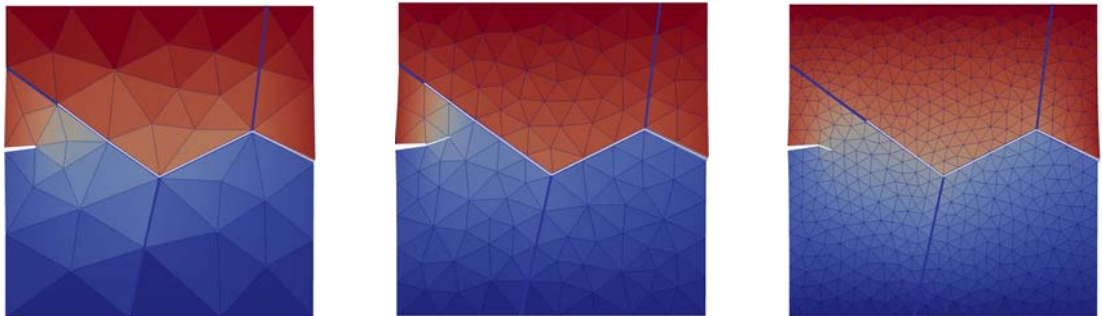
(d) $\sigma_{\text{peak}}^{\text{GB}} / \sigma_{\text{peak}}^{\text{grain}} = 0.40$ and 3 different mesh sizes

Figure 15. Contour plots of the vertical displacements at maximum load, for different mesh sizes (columns) and different ratios $\sigma_{\text{peak}}^{\text{GB}} / \sigma_{\text{peak}}^{\text{grain}}$ (rows from a to d).

## 5. Conclusions

A computational framework for the simulation of intergranular and transgranular cracking in polycrystalline Silicon solar cells has been proposed in the present work. To this aim, a specific image analysis procedure and a Matlab pre-processor have been developed to generate FE meshes with embedded interface elements. To avoid mesh dependency, the intrinsic CZM approach has been modified by adapting the initial stiffness of the linear part of the CZM by changing the mesh size.

Nonlinear finite element simulations on exemplary polycrystalline microstructures show a clear prevalence of transgranular cracking over intergranular fracture, unless very weak grain boundaries are present. These numerical results are consistent with the experimental observation in [6]. This pinpoints the necessity of simulating transgranular cracking in real solar cells. Clearly, the computational cost and convergence problems due to the very brittle nature of Silicon are enhanced. Indeed, further research is envisaged for the development of suitable techniques computationally affordable to simulate transgranular cracking in polycrystalline solar cells on a larger scale. Moreover, the presence of fingers and busbars, here accounted for as a global toughening effect, could also be properly modeled as a discrete reinforcement. Identification of CZM parameters is also expected to be crucial for reproducing the experimental stress-displacement curves in closer detail.

**Acknowledgements**

The research leading to these results has received funding from the European Research Council under the European Union's Seventh Framework Programme (FP/2007–2013)/ERC Grant Agreement No. 306622 (ERC Starting Grant "Multi-field and Multi-scale Computational Approach to Design and Durability of PhotoVoltaic Modules" – CA2PVM). The support of the Italian Ministry of Education, University and Research to the Project FIRB 2010 Future in Research "Structural mechanics models for renewable energy applications" (RBFR107AKG) is also gratefully acknowledged.